\documentclass[aps,prl,twocolumn,groupedaddress,showpacs,floatfix,superscriptaddress]{revtex4-1}
\usepackage{epsfig}
\usepackage{amsmath,amssymb}
\usepackage{graphicx}
\usepackage[dvipsnames,usenames]{color}
\usepackage[normalem]{ulem}
\usepackage{soul} % to use \st for strikeout 
\emergencystretch=\maxdimen
\begin{document}

\title{Charge Order in the Holstein Model
on a Honeycomb Lattice}
\author{Y.-X. Zhang}
\affiliation{Department of Physics, University of California, 
Davis, CA 95616,USA}
\email{zyxzhang@ucdavis.edu}
\author{W.-T. Chiu}
\affiliation{Department of Physics, University of California, 
Davis, CA 95616,USA}
\author{N.C. Costa} 
%% \email{natanael@if.ufrj.br}
%% \email{natanael.c.costa@gmail.com}
\affiliation{Instituto de F\'isica, Universidade Federal do Rio de Janeiro
Cx.P. 68.528, 21941-972 Rio de Janeiro RJ, Brazil}
\author{G.G. Batrouni}
\affiliation{Universit\'e C\^ote d'Azur, INPHYNI, CNRS, 0600 Nice, France}
\affiliation{MajuLab, CNRS-UCA-SU-NUS-NTU International Joint Research
  Unit, 117542 Singapore}
\affiliation{Centre for Quantum Technologies, National University of
  Singapore, 2 Science Drive 3, 117542 Singapore}
\affiliation{Department of Physics, National University of Singapore, 2
  Science Drive 3, 117542 Singapore}
\affiliation{Beijing Computational Science Research Center, Beijing,
100193, China}
\author{R.T. Scalettar}
\affiliation{Department of Physics, University of California, 
Davis, CA 95616,USA}

\begin{abstract}
The effect of electron-electron interactions on Dirac fermions, and
the possibility of an intervening spin-liquid phase between the
semimetal and antiferromagnetic (AF) regimes, has been a focus of
intense quantum simulation effort over the last five years.  We use
determinant quantum Monte Carlo (DQMC) simulations to study the Holstein model on
a honeycomb lattice and explore the role of electron-{\it phonon}
interactions on Dirac fermions.  We show that they give rise to charge-density-wave (CDW) order, and present evidence that this occurs only
above a finite critical interaction strength.  We evaluate the
temperature for the transition into the CDW which, unlike the AF
transition, can occur at finite values owing to the discrete nature of
the broken symmetry.
\end{abstract}

%\date{\today}

\maketitle

%%%%%%%%%%%%%%%%%%%%%%%%%%%%%%%%%%%%%%%%%%%%%%%%%%%%%%%%%%%%%%%%%%
%% \section{Introduction}
%%%%%%%%%%%%%%%%%%%%%%%%%%%%%%%%%%%%%%%%%%%%%%%%%%%%%%%%%%%%%%%%%%

\noindent
\underbar{Introduction:} The synthesis of graphene, {\it i.e.}~single
layers of carbon atoms in a hexagonal lattice, in 2004, has led to a
remarkable body of subsequent work\cite{geim07,blackschaffer14}.  One
of the key elements of interest has been the Dirac dispersion relation
of free electrons in this geometry, allowing the exploration of
aspects of relativistic quantum mechanics in a conventional solid.
%% including the possibility of unusual, topological superconducting 
%% phases\cite{blackschaffer07,nandkishore12,honerkamp08,wang12,kiesel12,pathak10,ma11,gu13,jiang14,xu16}.
``Dirac point engineering" has also become a big theme of investigation of
fermions confined in hexagonal optical lattices\cite{wunsch08}.

It has been natural to ask what the effects of electron-electron
interactions are on this unusual noninteracting dispersion relation.
Early quantum Monte Carlo (QMC) simulations and series expansion investigations of
the Hubbard model on a honeycomb lattice found a critical value of the
on-site repulsion $U_c \sim 4t$ for the onset of antiferromagnetic
(AF) order at half-filling\cite{paiva05}.  This stood in contrast to
the extensively studied square lattice geometry for which the perfect
Fermi surface nesting and the van Hove singularity of the density of
states (DOS) imply $U_c=0$.  Subsequent QMC studies refined this value
to $U_c \sim 3.87$ and suggested the possibility that a gapped,
spin-liquid (resonating valence bond) phase exists between the weak coupling semimetal and strong coupling AF regimes\cite{meng10}, a
conclusion further explored in the strong coupling (Heisenberg)
limit\cite{clark11}.  Yet more recent work challenged this scenario,
and pointed instead to a conventional, continuous quantum phase
transition (QPT) between the semimetal and AF
insulator\cite{sorella12,assaad13,otsuka16}. Equally interesting is
the possibility of unusual, topological superconducting phases arising
from these spin
fluctuations\cite{blackschaffer07,nandkishore12,honerkamp08,wang12,kiesel12,pathak10,ma11,gu13,jiang14,xu16}.

Graphene itself is, in fact, only moderately correlated.  First
principles calculations of the on-site Hubbard $U$ yield $U_{00} \sim
9.3$ eV\cite{wehling11}, with a nearest neighbor hopping $t \sim 2.8$
eV, so that $U/t \sim 3.3$ is rather close (and slightly below) $U_c$.
Longer range $U_{01}$ interactions can lead to a rich phase diagram
including charge ordered phases\cite{herbut06,honerkamp08}, especially
in the semimetal phase where the Coulomb interaction is unscreened.
Charge ordering may also arise when electron-phonon coupling (EPC) is taken into account\citep{gruner88,gruner94}.  Indeed, considering such
coupling would allow an exploration of the effect of other sorts of
interactions on the Dirac fermions of graphene, complementing the
extensive existing literature on electron-electron repulsion.

There are a number of fundamental differences between the two types of
correlations.  Most significantly, the continuous symmetry of the
Hubbard interaction, and the AF order parameter, preclude a
finite 2D temperature transition.  Therefore the focus is instead
on quantum phase transitions.  On the other hand, in the Holstein case
the charge-density-wave (CDW) order has a one-component order
parameter, leading to a transition that breaks a {\it discrete}
symmetry and, consequently, a finite critical temperature (in
the Ising universality class).  Precise QMC values of $T_c$ on a
square lattice were only quite recently
obtained\cite{costa17,weber18,costa18}. 
These build on earlier QMC studies of CDW physics in the Holstein model
\cite{marsiglio90,vekic92}, 
and introduce an exact treatment
of fluctuations into earlier mean-field calculations\cite{zheng97}.

{\it In this paper we explore the effect of electron-phonon, rather
  than electron-electron, interactions, on the properties of Dirac
  fermions, through QMC simulations of the Holstein
  model}\cite{holstein59} {\it on a honeycomb lattice.}  We use the charge
structure factor, compressibility, and Binder ratio to evaluate the
critical transition temperatures and EPC, leading to a determination of the phase diagram of the model.  Taken together, these results provide
considerable initial insight into the nature of the CDW transition for
Dirac fermions coupled to phonons.

\begin{figure}[t]
\includegraphics[scale=0.24]{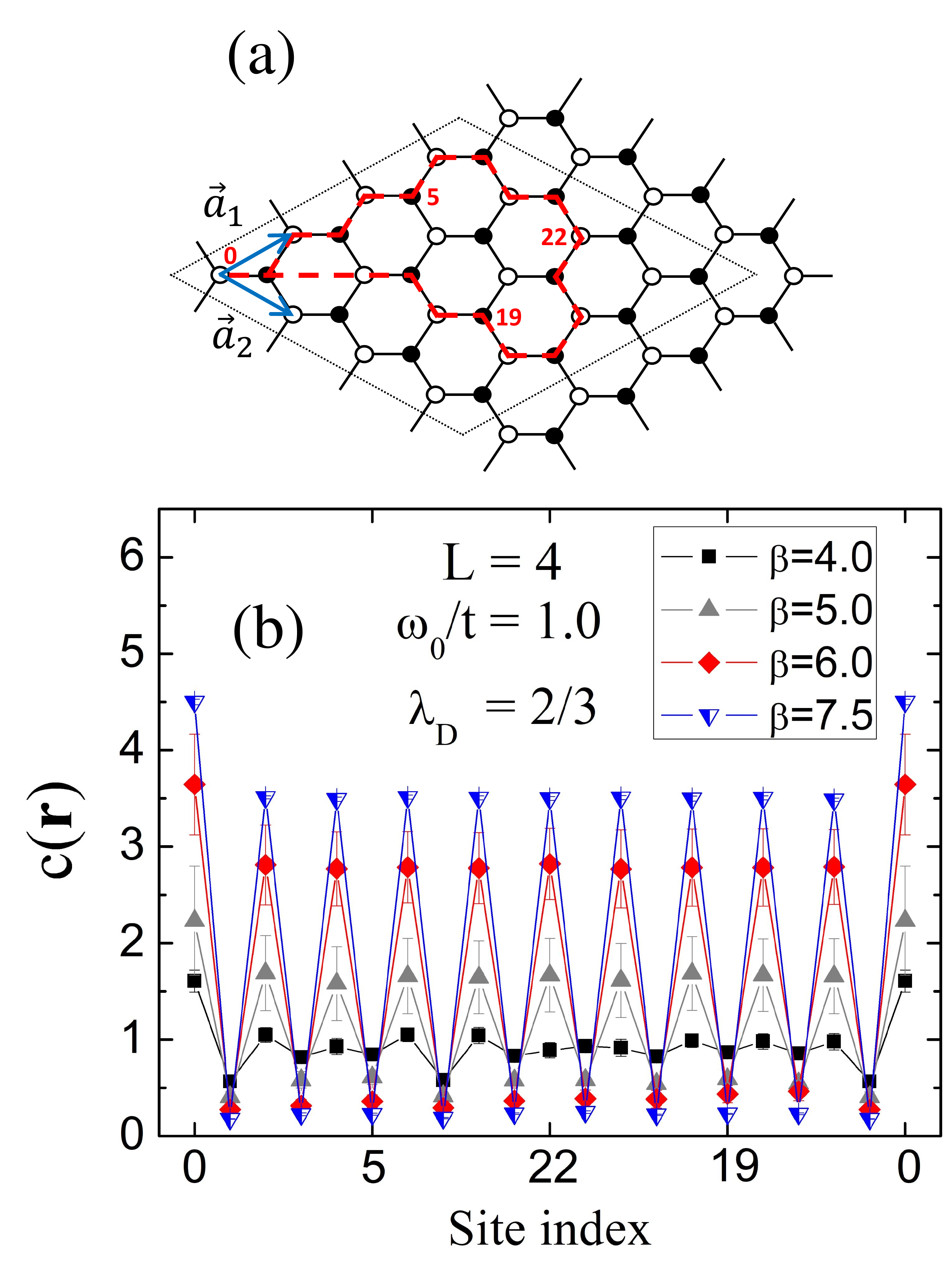} %
\caption{ (a) A $4 \times 4$ honeycomb lattice, with the trajectory (red
dashed line) corresponding to the horizontal axis of (b), which shows
charge correlations $c({\bf r})$ at $\lambda_D=2/3$, $\omega_0=1$, and
several temperatures.  Here, and in all subsequent figures, when not
shown, error bars are smaller than the symbol size.}
\label{fig1} % Fig1
\end{figure}

\vskip0.03in \noindent \underbar{Model and Methodology:} The Holstein
model\cite{holstein59} describes conduction electrons locally coupled
to phonon degrees of freedom,
\begin{align} \label{eq:Holst_hamil}
\nonumber \mathcal{\hat H} = & -t \sum_{\langle \mathbf{i}, \mathbf{j}
  \rangle, \sigma} \big(\hat d^{\dagger}_{\mathbf{i} \sigma} \hat
d^{\phantom{\dagger}}_{\mathbf{j} \sigma} + {\rm h.c.} \big) - \mu
\sum_{\mathbf{i}, \sigma} \hat n_{\mathbf{i}, \sigma} \\ & +
\frac{1}{2} \sum_{ \mathbf{i} } \hat{P}^{2}_{\mathbf{i}} +
\frac{\omega_{\, 0}^{2}}{2} \sum_{ \mathbf{i} }
\hat{X}^{2}_{\mathbf{i}} + \lambda \sum_{\mathbf{i}, \sigma} \hat
n_{\mathbf{i}, \sigma} \hat{X}_{\mathbf{i}} \,\,,
\end{align}
where the sums on $\mathbf{i}$ run over a two-dimensional honeycomb
lattice (see Fig.\ref{fig1}\,(a)), with $\langle
\mathbf{i}, \mathbf{j} \rangle$ denoting nearest neighbors.
$d^{\dagger}_{\mathbf{i} \sigma}$ and $d_{\mathbf{i} \sigma}$ are
creation and annihilation operators of electrons with spin $\sigma$ at
a given site $\mathbf{i}$.  The first term on the right side of
Eq.\,\eqref{eq:Holst_hamil} corresponds to the hopping of electrons,
with chemical potential $\mu$ given by the second term.  The phonons
are local (dispersionless) quantum harmonic oscillators with frequency
$\omega_{0}$, described in the next two terms of
Eq.~\eqref{eq:Holst_hamil}.  The EPC is included in the final term.
The hopping integral ($t=1$) sets the energy scale, with bandwidth
$W=6\,t$ for the honeycomb geometry.

We use determinant quantum Monte Carlo (DQMC) simulations \cite{blankenbecler81} to
investigate the properties of Eq.\eqref{eq:Holst_hamil}.  Since the
fermionic operators appear only quadratically in the Hamiltonian, they
can be traced out, leaving an expression for the partition function
which is an integral over the space and imaginary time dependent phonon
field.  The integrand takes the form of the square of the determinant of
a matrix $M$ of dimension the spatial lattice size, as well as a
"bosonic" action\cite{creutz81} arising from the harmonic oscillator
terms in Eq.\eqref{eq:Holst_hamil}.  The square appears since the traces
over the up and down fermions are identical, which leads to a case where
the minus sign problem is absent for any electronic filling.

%% Nevertheless, here we focus our attention on the half-filled case,
Nevertheless, we focus on the half-filled case,
$\langle \hat n_{{\bf i},\sigma} \rangle=\frac{1}{2}$.  This gives us
access to the Dirac point where the 
DOS vanishes linearly.  It is also the density
for which CDW correlations are most pronounced.  It can be shown,
using an appropriate particle-hole transformation, 
that this filling occurs at 
$\mu = -\lambda^2/\omega_0^2$.  We analyze lattices with linear
sizes up to $L=8$ (128 sites).  By fixing the discretization mesh to
$\Delta\tau=1/20$, systematic Trotter errors become smaller than the
statistical ones from Monte Carlo sampling.  To facilitate the
discussion, and eventual comparisons with the square lattice case, we
introduce a dimensionless EPC: $\lambda_D = \lambda^2 / (\omega_0^2 \, W)$. 

Charge ordering is characterized by the charge-density correlation function,
\begin{eqnarray}
c({\bf r}) = \big\langle \, 
\big( \, n_{{\bf i}\uparrow} + n_{{\bf i}\downarrow} \, \big)
\big( \, n_{{\bf i+r}\uparrow} + n_{{\bf i+r}\downarrow} \, \big) 
\, \big\rangle,
\end{eqnarray}
and its Fourier transform, the CDW structure factor,
\begin{eqnarray}
%% S_{\rm cdw} = \frac{1}{N} \sum_{\bf r} (-1)^{\bf r} c({\bf r})
S_{\rm cdw} =  \sum_{\bf r} (-1)^{\bf r} c({\bf r})
\,,
\end{eqnarray}
The $-1$ phase accesses the staggered pattern of the charge ordering.
The long-range behavior is investigated by performing finite size
scaling, and by tracking the evolution of the insulating gap
in the CDW phase.

%% \begin{figure}[h!]

\begin{figure}[t]
\includegraphics[scale=0.34]{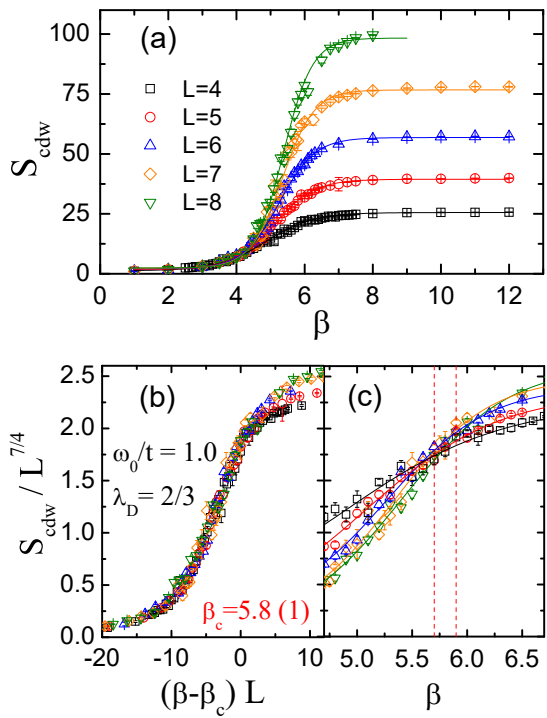} %
\caption{ (a) The charge structure factor as a function of $\beta$, for
different lattice sizes ($L=4$-$8$), and its (b) best data collapse,
with the 2D Ising critical exponents, which yields 
$\beta_c=5.8$.   
(c) The crossing
plot for $S_{\rm cdw}/L^{\gamma/\nu}$,
with vertical dashed lines indicating the uncertainty in the critical temperature.
Here $\lambda_D=2/3$ and $\omega_0=1$.
}
\label{fig2} % Fig2
\end{figure}

\vskip0.03in \noindent
\underbar{Existence of CDW phase:}
We first consider the behavior of charge-density correlations when the
temperature $T=\beta^{-1}$ is lowered.  Figure \ref{fig1}\,(b) displays
$c({\bf r})$ along the real space path of Fig. 1(a), for
$\lambda_D=2/3$, $\omega_0=1$ and several inverse temperatures $\beta$.
When $T$ is high ($\beta = 4$), we find $c({\bf r}) \approx \rho^2=1$, where $\rho$ is the density,
indicating an absence of long-range order.  However, an enhancement of
charge correlations starts to appear at $\beta = 5$, with the emergence
of a staggered pattern, which is even more pronounced at lower
$T$, $\beta = 6$ and $7.5$.  This temperature evolution of
real space charge correlations suggests a transition into a CDW phase.

A more compelling demonstration of long-range ordering (LRO) is
provided by Fig.\,\ref{fig2}\,(a), which exhibits the structure factor
$S_{\rm cdw}$ as a function of $\beta$, for different linear sizes
$L$.  In the disordered phase at high $T$, $c({\bf r})$ is
short-ranged and, consequently, $S_{\rm cdw}$ is independent of
lattice size $L$.  The emergence of a lattice size dependence of
$S_{\rm cdw}$, and, ultimately, its saturation at a value not far
from $N = 2L^2$, signals the onset temperature of LRO, and a
correlation length approaching the lattice size.  Figure
\ref{fig2}\,(a) shows that a change between these two behaviors occurs
around $\beta \sim 5-6$, giving an initial, rough estimate of
$\beta_c$.  The ground state is obtained for $\beta \gtrsim 8$; for
larger values, the density correlations no longer change.  The precise
determination of the critical temperature $T_c$ is accomplished by
performing finite size scaling of these data, using the 2D Ising
critical exponents $\gamma=7/4$ and $\nu=1$, as displayed in
Fig. 2(b).
The best data collapse occurs at
$\beta_c = 5.8\,(1)$, consistent with the crossing of
$S_{\rm cdw}/L^{\gamma/\nu}$ presented in Fig. 2(c), and
also supported by the crossing in the Binder cumulants (see Supplemental Material \cite{sm,binder81}).
%The best data collapse occurs at
%\textcolor{red}{$\beta_c = 5.8 \pm 0.1$, consistent with the crossing
%in the inset.}  \textcolor{blue}{This $T_c$ is also supported by
%data for the Binder crossing.  See Supplemental Materials.}
$T_c$ for the honeycomb lattice
is of the same order as that for the square lattice.  For the latter
at $\omega_0=1$, $\beta_c$ ranges from $\beta_c
\sim 16.7$ at $\lambda_D = 0.15$ to $\beta_c \sim 5$ at $\lambda_D =
0.27$ \cite{weber18}, and $\beta_c \sim 6.0$ at $\lambda_D =
0.25$ \cite{costa18,batrouni18}.  

For the range of EPC shown in
Ref.\,\onlinecite{weber18}, $\beta_c$ steadily decreases with
increasing $\lambda_D$.  A dynamical mean-field theory
approach \cite{blawid01,Freericks93} found that there is a minimal $\beta_c$
(maximum in $T_c$) for an optimal coupling strength.  This
non-monotonicity is also present in the repulsive half-filled 3D
Hubbard model; the AF $\beta_{\rm Neel}$ has a minimum at intermediate
$U$.  We return to this issue in what follows.

\begin{figure}[t]
\includegraphics[height=2.5in,width=3.5in]{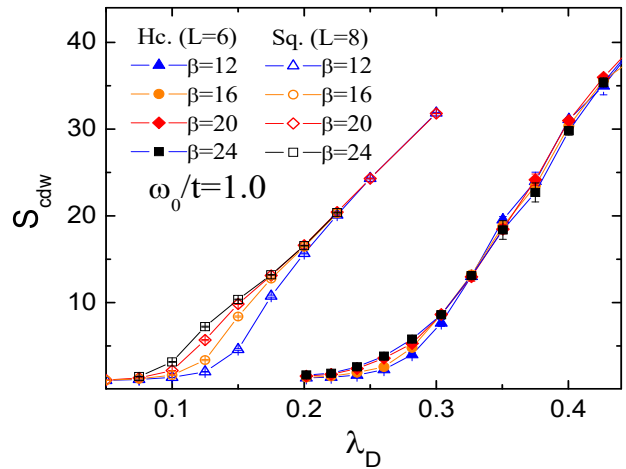} %
\caption{ CDW structure factor $S_{\rm cdw}$ as a function of
  dimensionless coupling $\lambda_D$.  $S_{\rm cdw}$ becomes small for
  $\lambda_D \lesssim 0.25$.  For the square lattice,
  $S_{\rm cdw}$ is large to much smaller values of $\lambda_D$.  In
  addition, for the honeycomb (Hc.) lattice $S_{\rm cdw}$ does not change
  for the two lowest temperatures, whereas $S_{\rm cdw}$ continues to
  grow at weak coupling for the square (Sq.) lattice.
}
\label{fig3} % Fig3
\end{figure}

\vskip0.03in \noindent \underbar{Finite Critical Coupling:} We
investigate next how charge correlations behave as a function of the EPC,
and, specifically the possibility that CDW does not occur below a
%% critical $\lambda_D$, as is known to be the case for the Hubbard model
critical interaction strength, as is known to be the case for the Hubbard model
on a honeycomb lattice.  This is a somewhat challenging question,
since at weak coupling one might expect $T_c \sim \omega_0 \,
e^{-1/\lambda_D}$ becomes small, necessitating a careful distinction
between the absence of a CDW transition and $T_c$ decreasing below the
simulation temperature.  Figure \ref{fig3} displays the CDW structure
factor as a function of $\lambda_D$ at different $T$, on square
(open symbols) and honeycomb (filled symbols) lattices, for similar
system sizes. The most noticeable feature is that
$S_{\rm cdw}$ appears to vanish for weak coupling, 
$\lambda_D \lesssim 0.25$, 
strongly suggesting a finite critical EPC for CDW order on
the honeycomb lattice.  This is a qualitatively reasonable consequence
of the vanishing DOS at half-filling, since having a finite DOS
is part of the Peierls' requirement for CDW
formation\cite{peierls55,gruner88,gruner94}.

To ensure this is not a finite $T$ effect, we contrast this behavior
of $S_{\rm cdw}$ with that of the square lattice, for which it is
believed that a CDW transition occurs at all nonzero $\lambda_D$ owing
to the divergence of the square lattice DOS\cite{weber18}.  We note
first that $S_{\rm cdw}$ remains large for the square lattice down to
values of $\lambda_D$ a factor of $2-3$ below those of the honeycomb
lattice.  In addition, there is a distinct difference in the $T$
dependence.  In the square lattice case, CDW correlations are enhanced
as $T$ is lowered.  The $S_{\rm cdw}$ curves shift systematically to
lower $\lambda_D$ as $\beta$ increases, consistent with order for all
nonzero $\lambda_D$.  On the other hand, $S_{\rm cdw}$ shows much less
$T$ dependence in the honeycomb case, 
with results from $\beta=12$ to
$20$ being almost identical (within error bars).

\begin{figure}[t]
\includegraphics[scale=0.30]{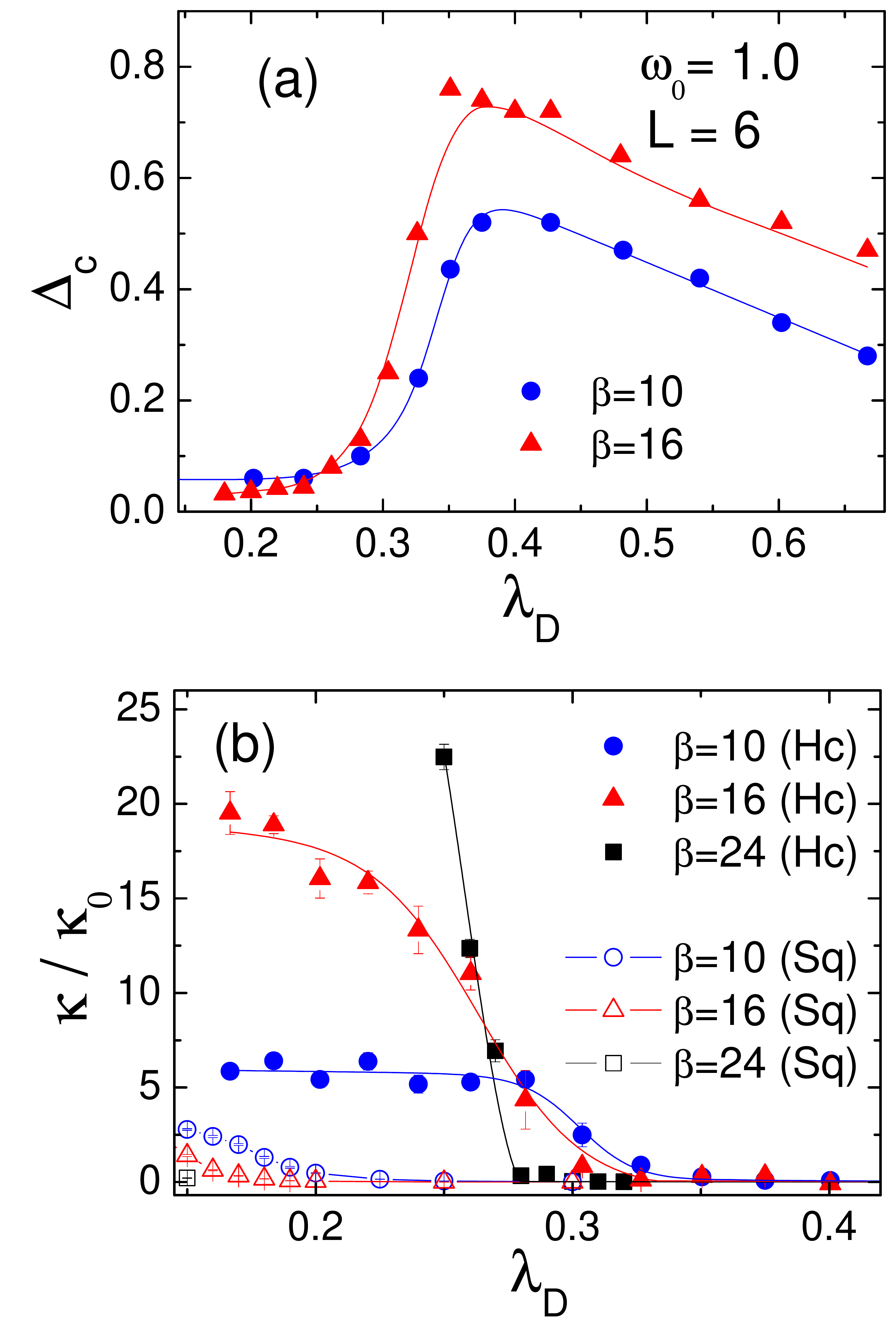} %
%% \caption{ (a) The electronic density as a function of the chemical
  %% potential for several couplings strength.  
  %% We have shifted the chemical potential so that half-filling occurs at
%% ${\tilde \mu}=\mu -\lambda^2/\omega_0^2=0$ for all curves.  
  %% }
  \caption{
  (a)  The charge gap $\Delta_c$ (see text) as a function of $\lambda_D$. 
 (b) The electronic compressibility
 $\kappa$ as a function of $\lambda_D$ for square (open symbols) and
 honeycomb (filled symbols) lattices with linear sizes $L=8$ and 6, respectively.  
 }
\label{fig4} 
\end{figure}

Further insight into the existence of a critical EPC is provided by
%% the evolution of the 
CDW gap, inferred from the 
plateau in $\rho(\mu)$ via
$\Delta_c \equiv \mu(\rho=1+x) - \mu(\rho=1-x)$.  Here we choose
$x=0.01$; other values of $x$ give qualitatively
similar results.
%% as a function of $\lambda_D$.  
Figure \ref{fig4}\,(a) displays $\Delta_c$
%% the charge gap, the plateau in $\rho(\mu)$,
for different $\lambda_D$ and fixed $\beta=10$ and $16$.  The gap has a
non-monotonic dependence on the EPC, with a maximum at $\lambda_D
\approx 0.43$.  For smaller EPCs the CDW gap is 
strongly suppressed.
%% first suppressed, 
%% $\lambda_D=0.33$, and finally is completely absent at
%% $\lambda_D=0.24$.  The absence of a gap is further confirmed by the
%% fact that the compressibility $\kappa=d\rho(\mu)/d\mu$ at half-filling
%% {\it increases} at lower temperatures ($\beta=16$).  
A crossing of the curves occurs 
at $\lambda_D \sim 0.27$  so that
$\Delta_c$ {\it decreases} as $T$ is lowered for
$\lambda_D \lesssim 0.27$, 
%% whereas the gap {\it decreases} as $T$ is lowered for
%% $\lambda_D \lesssim 0.27$, 
consistent with a critical EPC. 
The compressibility $\kappa=\partial\rho/\partial \mu$ is
presented as a function of $\lambda_D$ in Fig.~\ref{fig4}\,(b) for
honeycomb and square lattices at several $T$.  We have normalized by the noninteracting value $\kappa_{0}$ (evaluated in the thermodynamics
limit) to provide a comparison that eliminates trivial effects of the
DOS.  For the honeycomb lattice, $\kappa/\kappa_0$ shows a sharp
increase around $\lambda_D \sim 0.27 \pm 0.01$, consistent with the vanishing of
$S_{\rm cdw}$ in Fig.\,\ref{fig3}.  Furthermore, $\kappa/\kappa_0$
grows with $\beta$.  For the square lattice, $\kappa/\kappa_0$
vanishes down to much smaller $\lambda_D$, behaves more
smoothly at the lowest $T$, 
and is an order of magnitude smaller.
Its small residual value is a consequence of the exponentially divergence of
the CDW ordering temperature as $\lambda_D \rightarrow 0$.

\begin{figure}[t]
\includegraphics[scale=0.3]{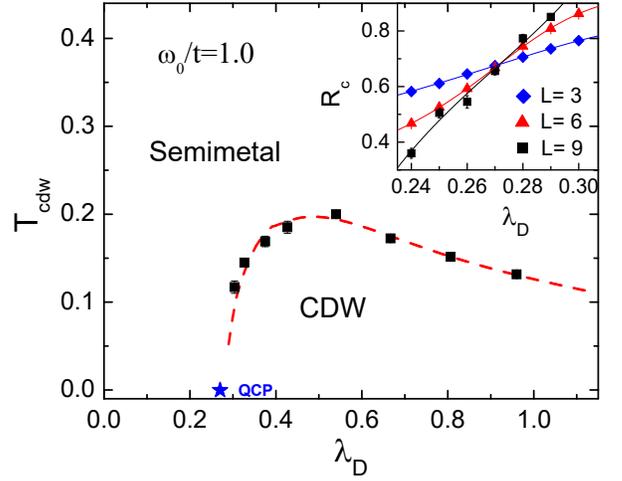} %
\caption{Critical temperature for the CDW transition in the honeycomb
  Holstein model inferred from finite size scaling analysis
  in Fig.~2.  
 The inset shows the crossing of the 
 invariant correlation ratio $R_c$ (see text),
 resulting in the indicated QCP, in good agreement with the value at which
 an extrapolated $T_c$ would vanish.
  %% The vertical grey bar is our estimate for
  %% $\lambda_{D,crit}$ below which $T_c$ vanishes.
  }
\label{fig5} % Fig5
\end{figure}

Finally, we have obtained $T_c$ for a range of $\lambda_D$ above the
critical EPC, yielding the phase diagram in Fig.\,\ref{fig5}.
$T_c$ decreases rapidly at $\lambda_D \approx 0.28$. 
The inset shows the crossing of the invariant correlation ratio $R_c$,
a quantity which is independent of lattice size at a quantum critical
point (QCP)(see Supplemental Material \cite{sm,binder81}).
$T_c$ exhibits a maximum at $\lambda_D \sim 0.4$-$0.5$, which lies
close to the coupling for which $\Delta_{\rm cdw}$ is greatest 
(Fig.\,\ref{fig4}).  The maximum in $T_c$ reflects a competition
between a growth with $\lambda_D$ as it induces CDW order with a
reduction as the EPC renormalizes the single electron mass, yielding a
heavy polaron
\cite{jeckelmann98,bonca99,romero99,ku02,kornilovitch98,hohenadler04,macridin04,goodvin06,sm}.
Unlike CDW order which arises directly
from intersite interactions, in the Holstein model it is produced by a
second order process: the lowering of the kinetic energy by virtual
hopping between doubly occupied and empty sites.  A mass
renormalization-driven reduction in this hopping lowers $T_c$.

\vskip0.03in \noindent \underbar{Conclusions:} In this paper we have
presented DQMC simulations of the Holstein model on a honeycomb
lattice.  The existence of long-range charge order was established
below a finite critical transition temperature in the range $T \sim
t/6$, for sufficiently large EPC.  
$T_c$ is similar for the square and honeycomb
lattices, despite the dramatic differences in their noninteracting
densities of states: diverging in the former case, and vanishing in
the latter.

Our data suggest that, as for the honeycomb Hubbard model,
\cite{paiva05,meng10,clark11,sorella12,assaad13,otsuka16}, the
vanishing non-interacting density of states of Dirac fermions gives
rise to a minimal value for $\lambda_D \sim (0.27 \pm 0.01)\,t$, only above which
does LRO occur.  Thus although the critical CDW transition
temperatures for the two geometries are similar {\it when order
  occurs}, the Dirac density of states does fundamentally alter the
phase diagram by introducing a weak coupling regime in which order is
absent.
%%  SEE PAGE 3 COLUMN 1 OF WEBER-HOHENADLER:
The 1D Holstein model is also known to have a metallic phase for
electron-phonon couplings below a critical value.
\cite{jeckelmann99,hohenadler17}

%% The half-filled 2D square lattice, on the other hand, 
%% is particularly special since it combines
%% a logarithmic divergence of the density of states at the
%% Fermi surface with perfect nesting.  This leads, in the
%% case of the Hubbard model, to an instability at any finite
%% $U>0$\cite{hirsch85},
%% and was also suggested to play a role in high superconducting
%% transition temperatures\cite{hirsch86,scalapino87}.

%% The investigation in this paper has uncovered both important
%% quantitative details concerning the effects of electron-phonon
%% interactions on Dirac fermions, and also confirms a qualitative 
%% distinction from the square lattice geometries which have been extensively
%% investigated for the case of the Hubbard Hamiltonian.

This initial study has focused on a simplified model.
The phonon spectra of graphene and graphitic materials
have been extensively 
%% studied\cite[and references cited therein]{karssemeijer11}
explored\cite{karssemeijer11} and, of course, are vastly more complex
than the optical phonon mode incorporated in the Holstein Hamiltonian.
However, as has been recently emphasized\cite{costa18}, including
realistic phonon dispersion relations is relatively straightforward in
QMC simulations, since the associated modifications affect only the
local bosonic portion of the action, and not the computationally
challenging fermionic determinants.  One important next step will be the
study of more complex phonon modes, and the types of electronic order
and phase transitions that they induce.  
Such investigations open the door to examining hexagonal CDW 
materials like the
%% NbSe$_2$, and 
transition metal dichalcogenides
\cite{Zhu2367,PhysRevB.89.235115,PhysRevB.89.165140,PhysRevB.57.13118}.
However, their layered structures add considerable challenges to descriptions
with simple models.
%% Another avenue is to examine
%% the physics away from half-filling, and the possibility of
%% superconducting transitions, including those of unusual symmetry.

{\it Note added.}---While preparing this manuscript, we learned of a
related investigation by Chen \textit{et al.} \cite{Chen18}.

%%%%%%%%%%%%%%%%%%%%%%%%%%%%%%%%%%%%%%%%%%%%%%%%%%%%%%%%%%%%%%%%%%
%% \section*{ACKNOWLEDGMENTS}
%%%%%%%%%%%%%%%%%%%%%%%%%%%%%%%%%%%%%%%%%%%%%%%%%%%%%%%%%%%%%%%%%%
\vskip0.03in \noindent
\underbar{Acknowledgements:}  The work of Y.-X.Z., W.-T.C. and R.T.S. 
was supported by the Department of Energy under Award No. DE-SC0014671.
G.G.B. is partially supported by the French government, through the 
UCAJEDI Investments in the Future project managed by the National 
Research Agency (ANR) with Reference No. ANR-15-IDEX-01.
N.C.C. was supported by the Brazilian funding agencies CAPES and CNPq.

%%%%%%%%%%%%%%%%%%%%%%%%%%%%%%%%%%%%%%%%%%%%%%%%%%%%%%%%%%%%%%%%%%%%%%%%
%%%
%%%%%     BIBLIOGRAPHY
%%%%%%%%%%%%%%%%%%%%%%%%%%%%%%%%%%%%%%%%%%%%%%%%%%%%%%%%%%%%%%%%%%%%%%%%%%%

\bibliography{bibyuxiholstein}
 
\end{document}

\beginsupplement

\title{Supplemental Material for \\
``Charge Order in the Holstein Model
on a Honeycomb Lattice"}
\maketitle
\date{\today}

\vskip0.10in \noindent

We present here additional measurements which support the conclusions of the main manuscript
and also provide complementary insight.

\vskip0.10in \noindent
\underbar{Qualitative Physics of the Holstein Model}

Electron-phonon interactions are generally expected to
give rise to pair formation. This is easily seen in the
Holstein model by considering the zero hopping ($t=0$) limit.  Tracing
out the phonon degrees of freedom yields an 
{\it attractive} electron-electron interaction 
$ U_{\rm eff}  n_{\uparrow} n_{\downarrow}$.
$U_{\rm eff}=-\lambda^2/\omega^2 $.
As the temperature is lowered, 
the pairs which are thus favored can either remain mobile
and hence condense into a superfluid state, or, at commensurate filling $\rho=1$ on a bipartite lattice, 
lock into an insulating pattern in 
which doubly occupied and empty sites occupy the two sublattices.
The alternation is favored by the same physics which 
leads to antiferromagnetic ordering in the repulsive Hubbard model.
If doubly occupied sites are adjacent, the Pauli principle
forbids hopping, while if a doubly occupied site is next to
an empty one, there is a second order {\it lowering} of the energy
$-t^2/U_{\rm eff}$.  
This is the CDW phase under consideration in this paper.

\vskip0.10in \noindent
\underbar{DQMC Algorithm}

Simulations were carried out with Determinant Quantum Monte 
Carlo (DQMC).  In this approach, no truncation of the phonon
Hilbert space is required, as is sometimes done in methods
which explicitly enumerate all occupation number states.
Instead, all sectors are explicitly included as the simulation
samples the (unrestricted) space and imaginary time 
dependent phonon coordinate. The inverse of the matrix $M$
which arises from tracing out the (quadratic) fermion degrees of 
freedom is the electron Greens function
$G(i,j)=-\langle \, c^{\phantom{\dagger}}(i)c^{\dagger}(j)\, \rangle$.
Averaging the diagonal elements yields the density, while
off diagonal elements give the kinetic energy.
By performing appropriate 
Wick contractions, expectation values of two particle
operators like the double occupancy and density structure
factor are obtained.
Thus, even though the
fermions have been integrated out, their correlation functions are
readily available.
Indeed, because $G$ is already 
needed to perform the phonon field update,
these measurements can be performed at very little
additional computational cost.
%% In the expression for the partition
%% function ${\cal Z} = e^{-\beta \hat H}$ the 
%% inverse temperature $\beta$ 
%% is discretized $\beta = L \Delta \tau$ and the full 
%% imaginary time evolution operator is expressed as a product 
%% $e^{-\beta \hat H} =
%% e^{-\Delta \tau \hat H} =
%% e^{-\Delta \tau \hat H} =
%% \cdots
%% e^{-\Delta \tau \hat H} }.
%% Complete sets of phonon position eigenstates are inserted, and

%%%%%%%%%%%%%%%%%%%%%%%%%%%%%%%%%%%%%%%%%%%%%%%%%%%%%%%%%%%%%%%%%%%%%%%%
%%%%%     Binder Ratio
%%%%%%%%%%%%%%%%%%%%%%%%%%%%%%%%%%%%%%%%%%%%%%%%%%%%%%%%%%%%%%%%%%%%%%%%

\vskip0.10in \noindent
\underbar{Binder Ratio:}

The Binder Ratio\cite{binder81} provides another means to determining critical points.  In Figure S1
we show results for
\begin{align}
B = \frac{\langle \,S_{\rm cdw}^2 \,\rangle}{\langle \,S_{\rm cdw} \, \rangle^2} \,,
\end{align}
which is the direct analog of the usual $\langle \, M^4 \, \rangle \, / \, \langle \, M^2 \, \rangle^2$
since $S_{\rm cdw}$ is already a density-density correlation.  See Eqs.~(2,3).
$B$ has a crossing in the range $5.8 < \beta_c < 6.2$, a range which overlaps
with the critical values inferred
from the $S_{\rm cdw}$ scaling collapse (Fig.~2b) and crossing (Fig.~2c)
plots.

We note that there is a subtlety in the meaning of the expectation values in Eq.~S1.
Our procedure is to measure $S_{\rm cdw}$ for a given field configuration,
via the appropriate Wick contractions of the
fermionic operators, and then square that {\it number}.  An alternate, and 
considerably more complicated procedure would involve the computation of
the full Wick contractions of the eight fermion operators in $S_{\rm cdw}^2$.
It is believed that the more simple procedure already captures the 
essential feature of the Binder procedure\cite{sandvik18}.
The consistency of our results with the analysis in the main paper confirm
this understanding.

\begin{figure}[t!]
\includegraphics[scale=0.30]{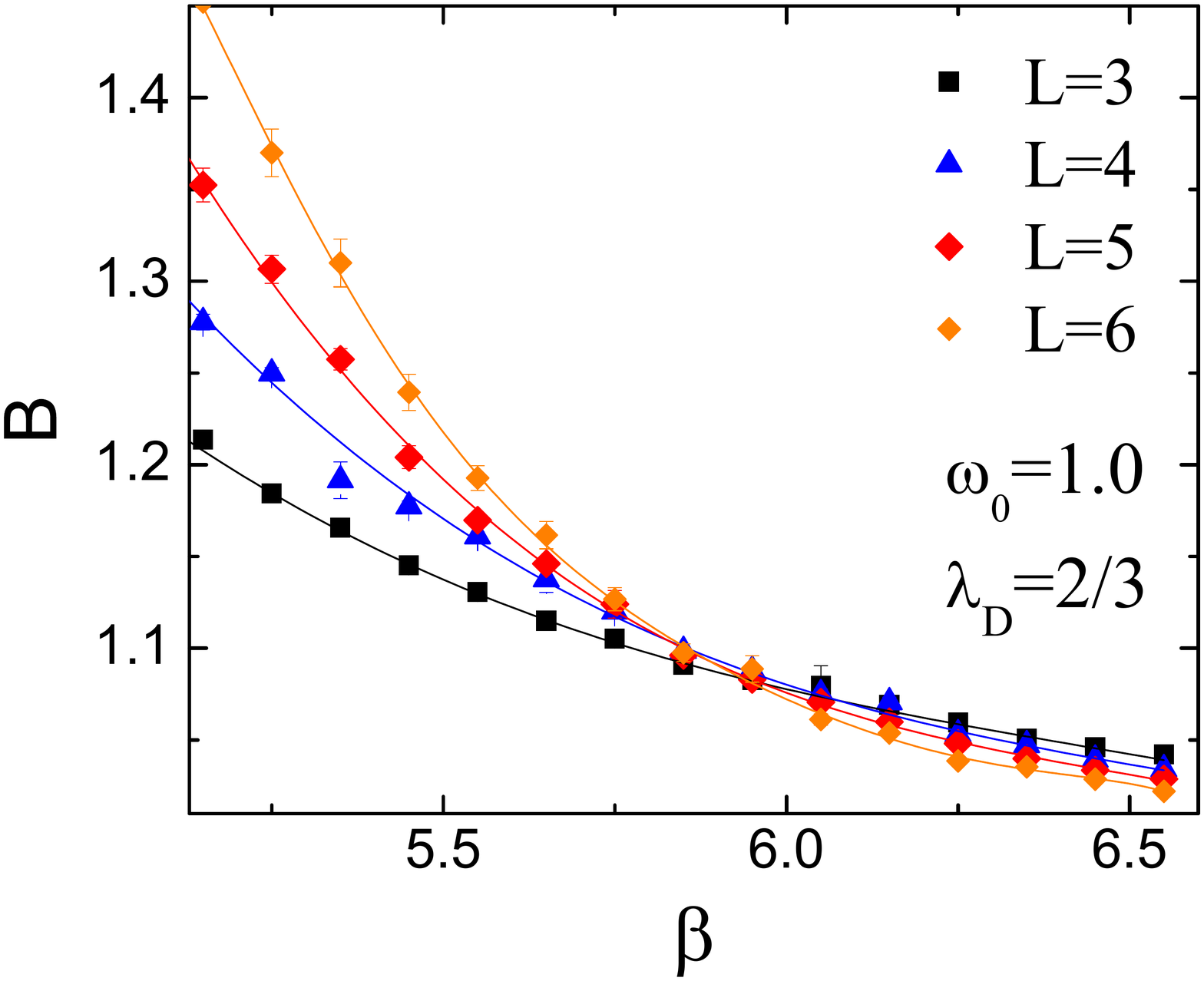}
\caption{Binder ratio for $\lambda_D=2/3$.  The curves for different lattice sizes cross at
$\beta_c \sim 5.8 - 6.2$, consistent with our estimate of the critical temperature
from the finite size scaling of the charge structure factor (Fig.~2b).
}
\end{figure}

%%%%%%%%%%%%%%%%%%%%%%%%%%%%%%%%%%%%%%%%%%%%%%%%%%%%%%%%%%%%%%%%%%%%%%%%
%%%%%     Electron Kinetic Energy
%%%%%%%%%%%%%%%%%%%%%%%%%%%%%%%%%%%%%%%%%%%%%%%%%%%%%%%%%%%%%%%%%%%%%%%%

\vskip0.10in \noindent
\begin{figure}[t!]
\includegraphics[scale=0.36]{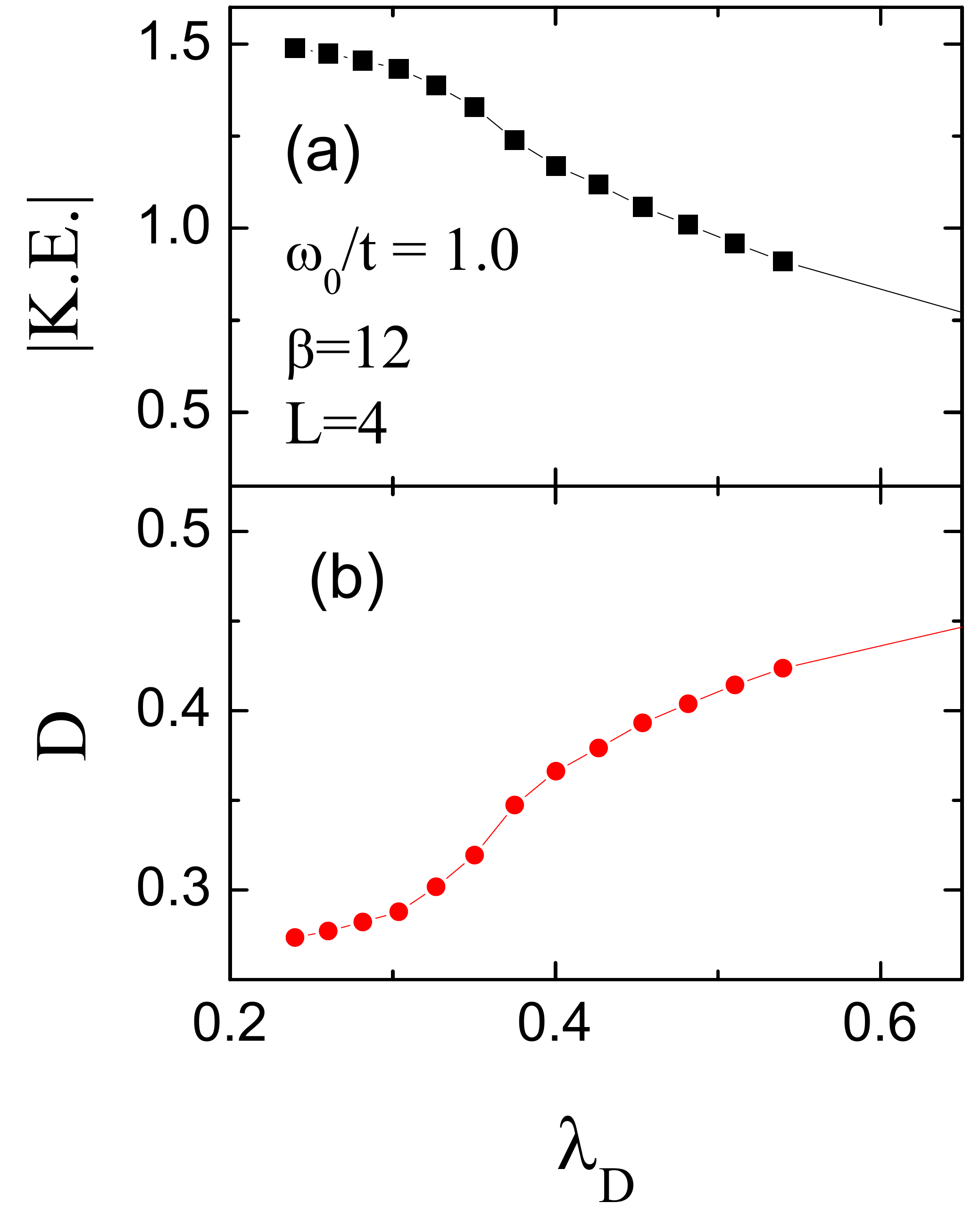}
\caption{
(a) The electron kinetic energy
and (b) double occupancy as a functions of $\lambda_D$
for an $L=4$ honeycomb lattice at $\beta=12$ and $\omega_0=1$.
%% The shaded grey area indicates our estimate for the critical
%% electron-phonon coupling below which there is no CDW order.
}
\end{figure}

\vskip0.10in \noindent
\underbar{Electron Kinetic Energy:}

We do not expect the fermion kinetic energy 
\begin{align}
KE = -t \, \sum_{\sigma}\langle \, 
c^{\dagger}_{{\bf i}\sigma}
\, c^{\phantom{\dagger}}_{{\bf j}\sigma}
+c^{\dagger}_{{\bf j}\sigma}
\, c^{\phantom{\dagger}}_{{\bf i}\sigma}
\, \rangle
\, ,
\end{align}
for near-neighbor sites ${\bf i,j}$,
to become zero in  
the CDW phase, since there are still local quantum fluctuations (hopping).
In fact, as pointed out in the main text, with only on-site interactions
such processes are required for the establishment of a CDW phase.  
Nevertheless, it should be reduced in magnitude as order is established.
Further, even in the absence of order, the fermions will become heavier 
with larger EPC and they evolve into a dressed `polaron', also suggesting
a reduced magnitude of kinetic energy.
This behavior is illustrated in Fig.~S2(top).
An analogous reduction in the electron kinetic energy is observed with
the increase of electron-electron interaction $U$ in the 2D repulsive
Hubbard model\cite{white89}.

%%%%%%%%%%%%%%%%%%%%%%%%%%%%%%%%%%%%%%%%%%%%%%%%%%%%%%%%%%%%%%%%%%%%%%%%
%%%%%     Double Occupancy
%%%%%%%%%%%%%%%%%%%%%%%%%%%%%%%%%%%%%%%%%%%%%%%%%%%%%%%%%%%%%%%%%%%%%%%%

\vskip0.10in \noindent
\underbar{Double Occupancy:}

In a perfect CDW phase at half-filling empty and doubly occupied sites alternate,
so that the double occupancy,
\begin{align}
D = \langle \, n_{{\bf i}\uparrow} n_{{\bf i}\downarrow} \, \rangle
\end{align}
goes to $D \rightarrow 0.5$.  In a completely uncorrelated phase, on other hand,
$D \rightarrow \langle n_{{\bf i}\uparrow}  \rangle \,
\langle  n_{{\bf i}\downarrow}   \rangle$
so that $D \rightarrow 0.25$ at half-filling.
The evolution of $D$ with  $\lambda_D$ is given in Fig.~S2(bottom) and properly exhibits
these two limits.  

The most rapid evolution of both $|KE|$ and $D$ in Fig.~S2 occurs
at $\lambda_D \sim 0.34-0.35$.  This is the same as the value for 
which $S_{\rm cdw}$ changes most quickly in Fig.~3 of the main text.
These quantities appear to become largely independent of $\lambda_D$ below the
critical value where CDW order no longer occurs.
%% and is also close to the coupling at which $\kappa$ vanishes in 
%% Fig.~4(b).

%%%%%%%%%%%%%%%%%%%%%%%%%%%%%%%%%%%%%%%%%%%%%%%%%%%%%%%%%%%%%%%%%%%%%%%%
%%%%%     Critical Exponents
%%%%%%%%%%%%%%%%%%%%%%%%%%%%%%%%%%%%%%%%%%%%%%%%%%%%%%%%%%%%%%%%%%%%%%%%

\vskip0.10in \noindent
\underbar{Critical Exponents:}

In the main paper we used the known 2D Ising exponents to do the finite size scaling analysis.  Here we verify the exponents
independently.  Specifically, we take the raw data for the CDW structure
factor, along with our estimate $\beta_c = 5.8$, and compute the scatter ${\cal S}$ of the scaled data $S_{\rm cdw}/L^{\gamma/\nu}$
for different lattice sizes $L=4,5,6,7,8$.  ${\cal S}$ is computed by taking data for all pairs $L_1, L_2$ and measuring
the root mean square difference of their scaled structure factors.
Fig.~S3 shows the result.  The minimum scatter is obtained at $\gamma/\nu \sim 1.85$.
We also investigate the critical exponents by performing data collapse using the Simplex algorithm, as described in Ref.\,\onlinecite{Melchert09} and references therein.
For the same DQMC data, this method provides $\gamma/\nu = 1.89(5)$ and $\nu = 1.0(1)$, for $-10 < (\beta-\beta_c) L^{\nu} < 10$, in line with the previous scatter analysis. 
The Simplex method also obtains the two exponents separately, as opposed
to just their ratio.
Within a shorter range, {\it e.g.}~$-6 < (\beta-\beta_c) L^{\nu} < 4$, one finds $\gamma/\nu = 1.8(1)$ and $\nu = 1.0(1)$.
This is in reasonable agreement with the 2D Ising values, 
%% which should be obtained in the thermodynamic limit, 
considering the finite lattice sizes employed.

\begin{figure}[t!]
\includegraphics[scale=0.30]{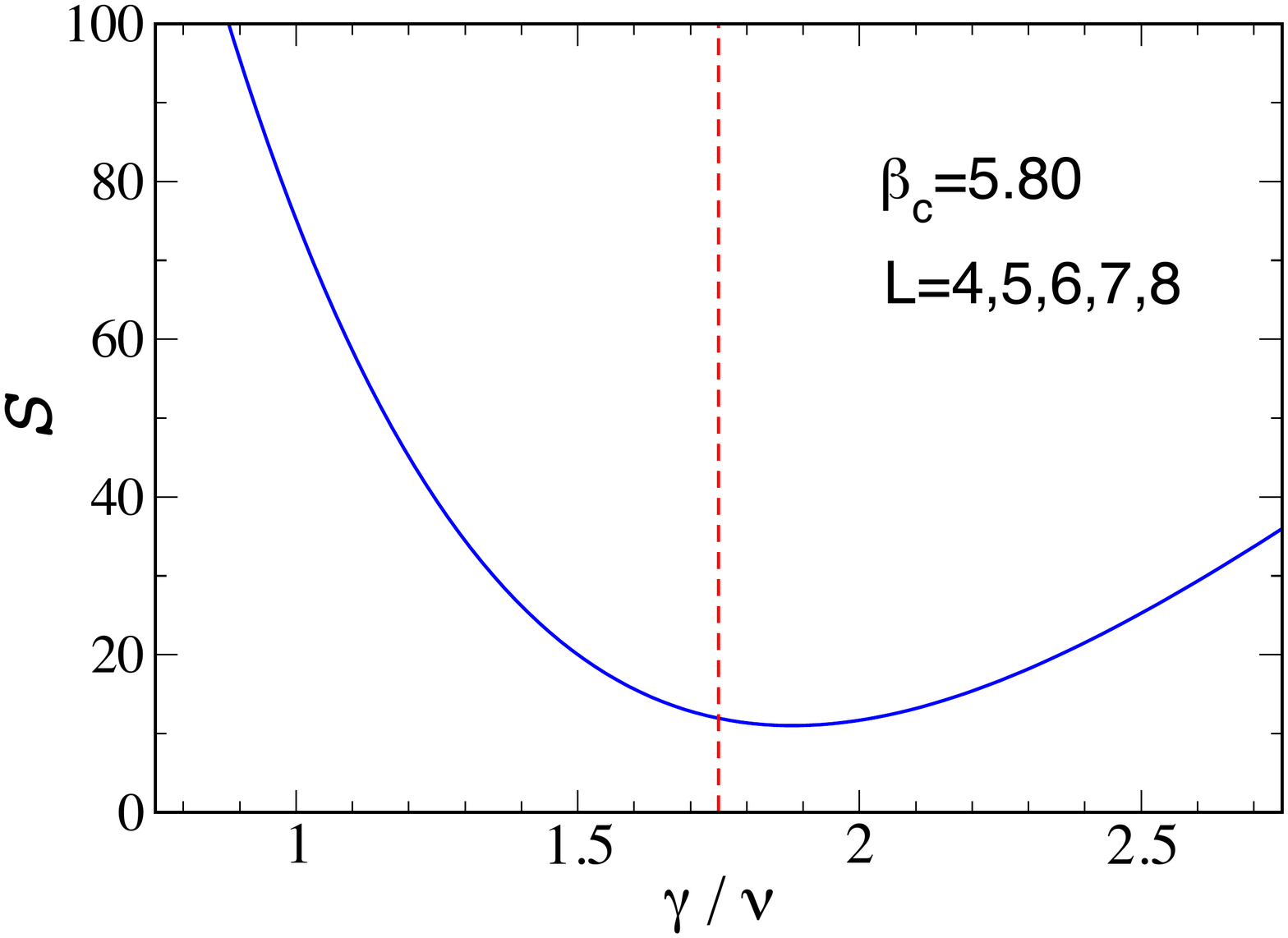}
\caption{
The scatter ${\cal S}$ (see text) of DQMC data for $S_{\rm cdw}/L^{\gamma/\nu}$ on different lattice
sizes as a function of scaling exponent $\gamma/\nu$.
The vertical dashed line is the 2D Ising value.
Data within a range $-10 < (\beta-\beta_c) L < 10$ were
used in computing ${\cal S}$.  The critical value was chosen
to be $\beta_c=5.8$.
}
\end{figure}

%%%%%%%%%%%%%%%%%%%%%%%%%%%%%%%%%%%%%%%%%%%%%%%%%%%%%%%%%%%%%%%%%%%%%%%%
%%%%%     CDW gap
%%%%%%%%%%%%%%%%%%%%%%%%%%%%%%%%%%%%%%%%%%%%%%%%%%%%%%%%%%%%%%%%%%%%%%%%

\vskip0.10in \noindent
\underbar{CDW gap:}

Further insight into the existence of a critical EPC is provided by
the evolution of the CDW gap as a function of $\lambda_D$, as already seen in Fig.~4(a) of the main text.  Fig.~S4 displays the charge gap by showing the
full plateau in $\rho(\tilde \mu)$,
for different $\lambda_D$ and fixed $\beta=10$. The gap has a
non-monotonic dependence on the EPC, with a maximum at $\lambda_D
\approx 0.43$.  For smaller EPCs the CDW gap is first suppressed, 
$\lambda_D=0.33$, and finally is completely absent at
$\lambda_D=0.24$.
The $\beta=16$ curve emphasizes that at weak coupling $\lambda_D=0.24$ the compressibility 
{\it grows} as $T$ is lowered. 

\begin{figure}[t!]
\includegraphics[scale=0.40]{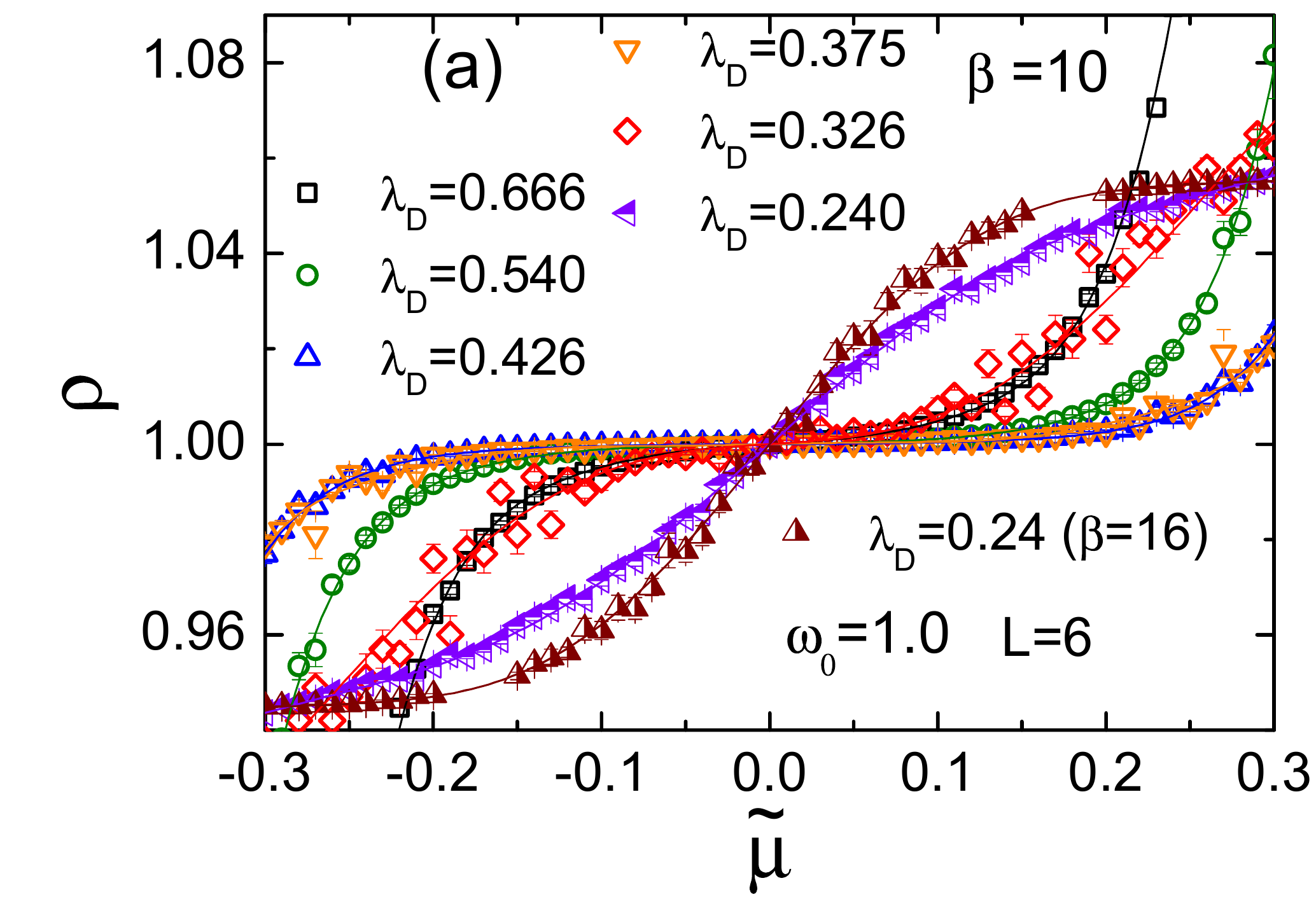}
\caption{
The electronic density as a function of the chemical
  potential for several couplings strength.  
  We have shifted the chemical potential so that half-filling occurs at
${\tilde \mu}=\mu -\lambda^2/\omega_0^2=0$ for all curves. 
}
\end{figure}

\vskip0.10in \noindent
\underbar{Invariant Correlation Length:}

The invariant correlation length $R_c$, 
\begin{align}
R_c \equiv 1 - \frac{S({\bf Q}+{\bf \delta q})}{S({\bf Q})}
\end{align}
shown as the inset to Fig.~5 of the main text,
measures the fall off of the charge structure factor
\begin{align}
S({\bf q})=\sum_{\bf r} e^{i {\bf q} \cdot {\bf r} } c({\bf r})
\end{align}
as ${\bf q}$ is shifted away from its CDW value at wavevector ${\bf Q}$
%% =(\pi,\pi)$ 
(Eq.~3 of main text).
$R_c$ is advantageous to consider as it has,
similar to the `Binder ratio',
more benign scaling corrections
than does the charge structure factor itself\cite{binder81}.